# Optical Vortex Beam Propagation in Turbulent Atmosphere

YuanYao Lin[1] and Shi-Cheng, Xiao[1]

[1] Department of Photonics,
National Sun Yat-Sen University, Kaohsiung City, 804

**Abstract**

Propagation of optical vortex beamlet (OVB) in turbulent atmospheric environment is studied theoretically by normalized models. Our simulation shows that optical vortex beamlet is superior to traditional Gaussian beam in terms of diffraction behavior of long haul propagation. Formation of vortex beam let is experimentally demonstrated by using a spatial light modulator (SLM), which achieve a power efficiency of more than 90% when the intrinsic loss in SLM is excluded.

## 1.Introduction

Optical Vortex Beam is a beam of electromagnetic wave at optical wave length that carries orbital angular momentum [1]. Optical vortex beam is an excellent candidate in applications based on laser free-space propagation. Optical vortex beams are more resistant to noises induced by turbulence in atmospheric environment and tend to heal themselves so that the angular momenta they carry are kept [2-4]. Therefore the deformation in the optical beam is minimized and pointing stability in the optical beam is improved. Because optical vortex beam has vanishing intensity at the center of the beam due to the phase singularity, it can induce self-guiding channel in a self-defocusing nonlinear system, such as thermal blooming in atmosphere. The propagation of high power optical vortex thus enjoys the self-guiding effect to achieve better propagation distance without a significant loss of optical power [4].

Optical vortex beams should be able to play an important role in applications related to defense technologies, specifically, remote sensing, free space communication, ranging and LIDARs. For this purpose, the development in high power optical vortex beams and the model to predict the optical vortex beam propagation in turbulent and nonlinear atmospheric environment is a crucial task.

This project aim at providing a practical evaluation of the figure of merit of the high power optical vortex beam propagation in turbulent and nonlinear atmospheric environment. The followings are to be investigated. First we build up (quasi-vortex) vortex laser beam by combining multiple beams of phase modulated lasers that are spatially managed at specific positions. Second we propose a phenomenal model that can simulate and predict the propagation of optical vortex beam in turbulent and nonlinear atmospheric environment. Based on the proposed model, a turbulent and nonlinear atmospheric environment simulator based on spatial light modulator is to be built to facilitate a test bed for vortex laser beam propagation in the future.

## 2. Propagation Model in Turbulence Air with Thermal nonlinearity

We consider the propagation of electromagnetic in atmospheric environment which is rich in turbulence and nonlinear effect [5-7].

$$i2k\frac{\partial u}{\partial z} + \nabla_\perp^2 u + 2k^2(\delta n_L + \delta n_T - i\alpha_l)u = 0 \quad (1)$$

Where $\nabla_\perp^2 \equiv \partial_x^2 + \partial_y^2$ is the transverse Laplacian, $\alpha_l$ is the total loss experienced by the electromagnetic wave including the absorption of the air and scattering losses. $\delta n_L$, and $\delta n_T$ are the perturbation in refractive index due to laser induced nonlinear effect and atmospheric turbulence. These perturbations follow the dynamical equation of air flows:

$$\left(\frac{\partial}{\partial t} + v \cdot \nabla - \chi\nabla^2\right)\delta n_L = \frac{-\alpha}{\rho c_p}\frac{dn}{dT}|U|^2 \quad (2)$$

in which $\nabla^2 \equiv \partial_x^2 + \partial_y^2 + \partial_z^2$ is the 3 dimensional Laplacian. For CW beams $\nabla^2 \approx \nabla_\perp^2$ and it require special care for pulsed operation if the longitudinal extent of the beam is comparable or even shorter than the transverse beam radius. $\chi$ is the thermal diffusivity given by $\chi = \frac{k_{th}}{\rho c_p}$, $k_{th}$ is the thermal conductivity, $\rho$ is the mass density and $c_p$ is the specific heat capacity of the air. $\alpha$ is the absorption coefficient of the air and $dn/dT$ is the temperature dependence of refractive index. The physical quantity of the parameters used are listed in Tab. 1.

| Symbol | Value | Physical Property |
|---|---|---|
| $\lambda$ | $0.633 \mu m$ | wavelength |
| $k = 2\pi/\lambda$ | | wavevector |
| $\chi$ | $1.9 \times 10^{-5}$ m²/s | thermal diffusivity |
| $1/(\rho c_p)$ | 770cm³ ⁰C/J | inverse of the product of mass density and specific heat capacity |
| $\alpha$ | 0.002km⁻¹ | absorption coefficient |
| $dn/dT$ | $2 \times 10^{-9}$ 1/⁰C | temperature dependence of refractive index |

Table 1. Listed are the symbols used in the equations and their physical values [5-7].

The refractive index perturbation associated with the atmospheric turbulence are well documented []. It can be estimated using various models, among which the modified Atmospheric model gives the most spectral characteristic compared to the experimental measurement reported previously [8]. Therefore modified Atmospheric model was adopted instead of Kolmogorov model.

$$\Phi_{PSD}(f) = 0.023 r_0^{-5/3} \left(f^2 + f_0^2\right)^{-11/6}$$
$$\times \left[1 + 1.802 \frac{f}{f_l} - 0.254 \left(\frac{f}{f_l}\right)^{7/6}\right]$$
$$\times \left(1 - e^{-\frac{f^2}{f_0^2}}\right) e^{-\frac{f^2}{f_l^2}}$$

(3)

where $f_1 = 3.3/l_0$ and $f_0 = 1/L_0$ in which $l_0$ and $L_0$ are the inner and outer scale of the turbulence. $C_n^2$ is the structure constant of the trubulence and $r_0$ is the Fried Parameter given by,

$$r_0 = 1.67 \left(k^2 \int_0^L C_n^2(z) dz\right)^{-3/5} \quad (4)$$

An example power spectral density distribution of the modified atmospheric model and Komogorov model are compared in Fig. 1 with inner and outer scales of 1cm and 177m, respectively.

For example, the atmospheric turbulence introduces phase aberration that corresponds to the perturbations in the refractive index changes of turbulence $\delta n_T = \Phi_T / k$ is illustrated in Fig. 2.

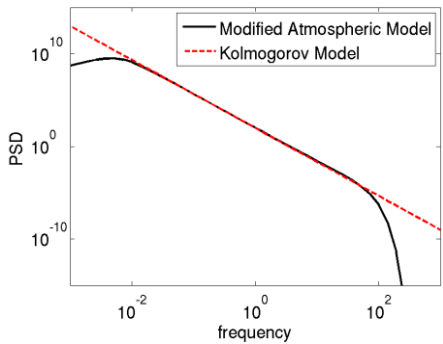

Fig. 1 Power spectral density distribution of the modified atmospheric model and Komogorov model are compared with $l_0$=1cm and $L_0$=175m, respectively.

(a)  (b)

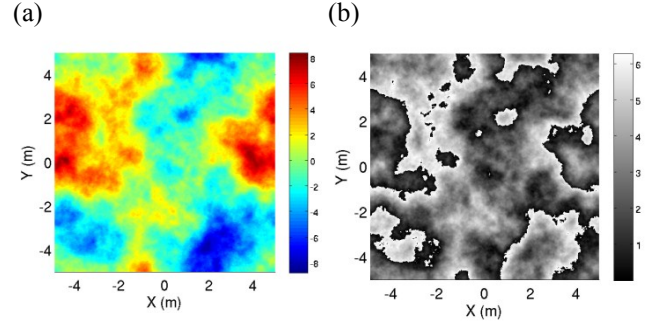

Fig 2. (a) Example of $\Phi_T$ constructed by random variables of normal distribution that is scaled by the power spectrum distribution of modified atmospheric model. (b): turbulence pattern prepared for spatial light modulator (SLM). $L_0$=175m and $l_0$=1cm and $r_0 \sim$ 1.1cm.

### 3. Normalization and Scaling of Nonlinear Propagation Equations

Although Eqs. (1), (2) and (3) are suffice to present the electromagnetic wave propagation in turbulent atmospheric environment, various parameters involved limit our understanding of the major role in the propagation effect. To get the most out of the model without brute forces, we normalize the equations so the propagation effect can be inferred and scaled. We normalize all the physical quantity into dimensionless quantities following the relations listed in table 2 in which the parameters $W_0$ is the radius of the laser beam, $L_N$ is the nonlinear length for the nonlinear phase to accumulate 1 radian. $T_0$ is the response time for thermal diffusion. $P_0$ is input power the laser beam.

| New Symbol | Original Symbol | Relation |
|---|---|---|
| $X, Y$ | $x, y$ | $X = W_0 x$, $Y = W_0 y$ |
| $Z$ | $z$ | $Z = z L_N$ |
| $\tau$ | $t$ | $\tau = T_0 t$ |
| $U$ | $u$ | $\sqrt{P_0} u$ |

Table 2 Relation between the original variable and the normalized quantities.

With such a dimensionless variable, the thermal dynamical equation gives,

$$\left[\frac{T_0}{W}\left(v_x \frac{\partial}{\partial X} + v_y \frac{\partial}{\partial Y}\right) + \frac{\chi T_0}{W^2}\left(\frac{\partial^2}{\partial X^2} + \frac{\partial^2}{\partial Y^2}\right)\right] \delta n_L$$
$$= \frac{\partial \delta n_L}{\partial \tau} - \frac{-\alpha}{\rho c_p} \frac{dn}{dT} T_0 |u|^2$$

(5)

in which the variation of refractive index structure is neglected. Here if we define a time scale for thermal equilibrium to reach $T_0 = W_0^2/\chi$ which is about $5.26 \times 10^4 W_0^2$ sec ($W_0$ is in unit of meter), then for a beam of diameter around 1m, it is way too long to reach a thermal equilibrium as is the case in most real propagation case. To emulate the real case in which the beam diameter is large, we neglect the thermal diffusion term and we rewrite the thermal dynamical equation

$$\frac{\partial \delta n_L}{\partial \tau} - \left[\frac{T_0}{W}\left(v_x\frac{\partial}{\partial X} + v_y\frac{\partial}{\partial Y}\right)\right]\delta n_L = \frac{-\alpha}{\rho c_p}\frac{dn}{dT}T_0|u|^2 \quad (6)$$

In the absence of wind, one has the near equilibrium state,

$$L_N = \frac{\rho c_p}{\alpha}\frac{1}{\frac{dn}{dT}k}(T_0 I_0)^{-1} = \frac{65.4}{T_0(sec)\cdot I_0(MW/m^2)} \quad (7)$$

For an optical beam of intensity 100kW/m² and the observation time around 1 second. $L_N \approx 654$m. In the presence of wind, the optical beam tends to deflect in the direction against the wind and the amount is proportional to wind speed. For current model, wind speed is neglected to get better scaling and the effect of turbulence.

$$i\frac{\partial U}{\partial Z} + \frac{L_N}{2kW^2}\left(\frac{\partial^2}{\partial X^2} + \frac{\partial^2}{\partial Y^2}\right)U - |U|^2 U + L_N k \delta n_T U = 0 \quad (8)$$

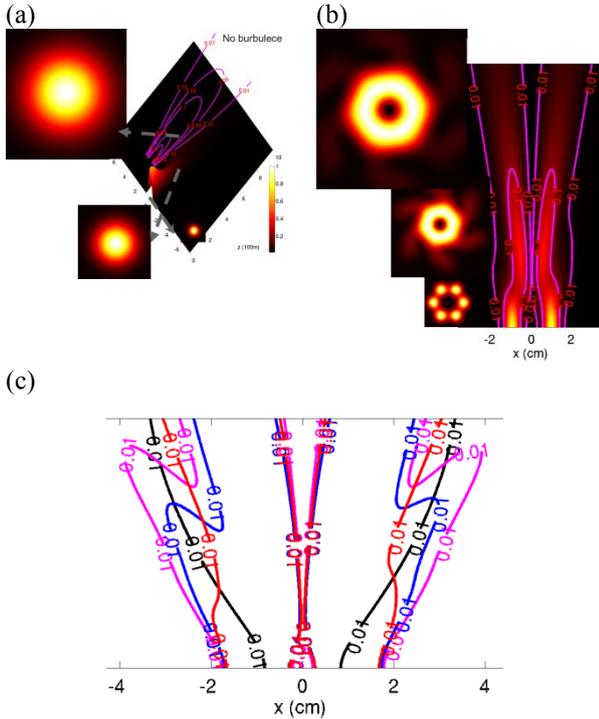

Fig 3. Without turbulence (a) Propagation of fundamental Gaussian beam. (b) Propagation of vortex beamlet. (c) Diffraction effect of optical beams. Fundamental beam, two humped beam, quadruple beamlet and sex-druple beamlets are in black, red, blue and magenta, respectively.

## 4. Simulation Result and Discussion

In the simulation, we choose beam diameter $W_0$ to be 1cm and a propagation distance to be 1 km. The total loss only counts the absorption that cause the temperature change of th air. For the turbulence effect we test for conditions with different Fried Parameters $r_0$ to indicate the degree of turbulence. Three different $r_0$ (2.8mm, 0.16mm and 0.07mm) are used to illustrate the strength of turbulence relative to the input beam radius. The nonlinear length $L_N \approx 654$m with optical intensity of 100kW/m². The topological charge of the vortex beamlet is kept 1, i.e. a $2\pi$ phase modulation in azimuthal direction. It is shown in Fig. 3 that without turbulence effect, one sees that vortex beamlet give little diffraction effect due to the thermal blooming effect that cause negative lens effect, which is also justified by previous literatures[2-4].

Under the effect of turbulence, simulation result shown in Fig. 4 reveals that the beam structure remain in good quality after a propagation of 1km. In specific, for the proposed sex-druple beamlet, the diffraction effect is preserved for various turbulence conditions as shown in Fig. 5. It means that the vortex beam structure has better resistance to the turbulence than fundamental Gaussian beam as can understood by comparing to the calculated beam width of second moment of the intensity distribution. Vortex beams experiences small beam cross section than the fundamental Gaussian beam in a long term propagation. Vortex beamlet are generated experimentally and the trubulence phase mask is demonstrated in Fig.6. for future investigations.

## 5. Conclusion

We formulate the model to analyze the propagation of optical vortex beam in turbulent environment. The model is normalized to exploit the scalability. Simulation based on the proposed model shows that optical vortex beamlet is more resistant to turbulence effect and thermal blooming. The generation of beam let and turbulence pattern are experimentally demonstrated. This research shows that optical vortex beamlet is promising for long-haul beaming applications.

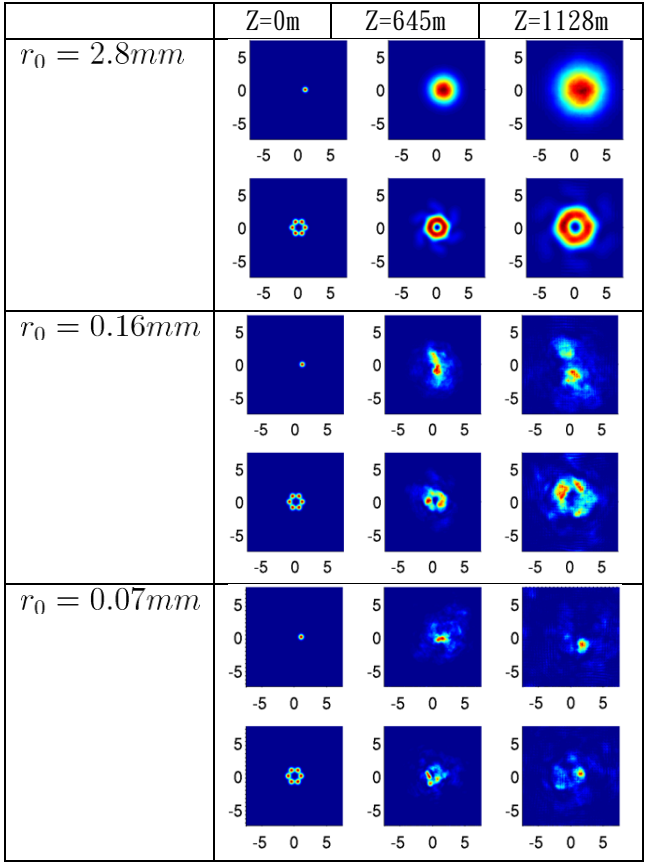

Fig 4. Propagation of 6-lobed beamlet under thermal blooming effect and with turbulence effect.

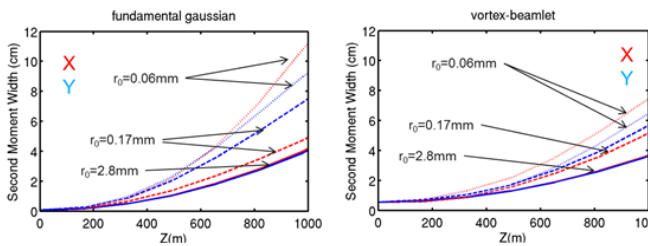

Fig 4. Beam width of the fundamental and vortex propagation through turbulent air.

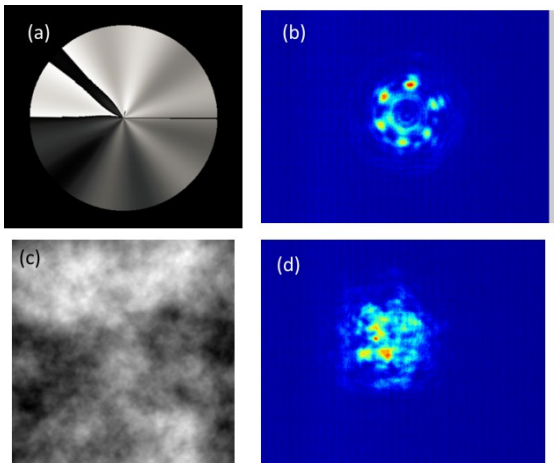

Fig. 6 Experimental demonstration of (a) phase modulation pattern in SLM for generating (b) beamlets (c) phase pattern on SLM to generate trubulence pattern with fried parameter of 0.1mm. (d) propagation of fundamental Gaussian modulated by the turbulence pattern in SLM.

## 6. Acknowledgement
The authors thank the support of National Chung-Shan Institute of Science and Technology.